# Cascade kernels for models of the turbulent fluid cascade

Miriam A. Forman

*Department of Physics and Astronomy, State University of New York at Stony Brook, NY11794-3800*

This paper presents calculations of the Castaing (Physica D, 46, 177, 1990) cascade kernels for five well-known models of the turbulent cascade and demonstrates that these kernels provide conceptually simple and direct descriptions of turbulence models equivalent to their multifractal spectra. The problem with a log-normal model seems to be that its Castaing kernel predicts a small but non-zero probability that finite-amplitude turbulence will decay to arbitrarily large turbulent amplitudes at smaller scales. Kernels for log-Poisson models do not have this pathology. It is shown how kernels evolve when the cascade model itself varies with scale. Future studies of other kernels compatible with and even dictated by detailed fluid physics may assist understanding of the turbulent cascade.

Castaing scaling [1,2,3,4] describes the cascade of turbulent energy in a fluid from scale L to smaller scale L' by relating the probability distribution functions (pdfs) of the velocity increments $\delta V$ at the different scales with a cascade kernel $G(u;L',L)$ such that

$$P_{L'}(\delta V) = \int_{-\infty}^{+\infty} G(u;L',L) e^{-u} P_L\left(e^{-u}\delta V\right) du. \qquad (1)$$

The kernel $G(u;L',L)$ represents the probability that turbulence of amplitude $e^{-u}\delta V$ on scale L, cascades to turbulence of amplitude $\delta V$ on scale L'. Castaing scaling is linear in $\delta V$, in the sense that the same $G(u)$ relates $ae^{-u}\delta V$ at the larger scale to $a\delta V$ at the smaller scale, for any $-\infty < a < \infty$. Looking at such kernels seems a good way to concretely visualize what is actually going on in the cascade. Multifractality and intermittency appear in the cascade when the kernel $G(u)$ is anything other than a delta-function. If both $G(u)$ and $P_L(u)$ are Gaussian functions of their arguments, $P_{L'}(u)$ will be a winged "Castaing distribution" characterized by the mean and width of $G(u)$. Such distributions can be conceptually useful (Sorriso-Valvo, et al. [5]). However, neither $G(u)$ nor $P_L(u)$ can be precisely Gaussian; if they were, the scaling would be log-normal, and the third moment of the parallel velocity increment would be zero, both of which are refuted fundamentally by Frisch [6]. The Castaing pdf is only approximate and should be used with care. This paper deals only with Castaing *scaling*.

Following equation 1, the scaling properties of all the moments of the pdf arise from the scaling of $G(u;L,L)$. In particular, since the moment of order q at scale L' is

$$S(q,L') = \int_{-\infty}^{+\infty} d(\delta V)(\delta V)^q \int_{-\infty}^{\infty} G(u;L',L)e^{-u} P_L(e^{-u}\delta V) du \qquad (2a)$$

$$= \left[\int_{-\infty}^{\infty} G(u;L',L)e^{qu} du\right] S(q,L), \qquad (2b)$$

all of the dependence on L' (in other words, the scaling) is contained in G(u:L',L).

Scaling is most often described with $\zeta(q)$, the logarithmic derivative of S(q, L') with respect to L' (see e.g. [6]). $\zeta(q)$ is called the exponent of the structure function. The moments of any actual data set exist for q = any non-negative integer. As an example, Kolmogorov's 4/5 law states that the third moment of the longitudinal velocity increment is negative, but scales as L'/L [6]. However, since the *scaling is completely subsumed into the kernel G, and does not involve the pdf at large scale*, the scaling of the moments of the absolute value of $\delta V$ is the same as the scaling of the moments of $\delta V$ for integer q $\geq$ 0. The odd moments of the absolute values are not the same as the odd moments of $\delta V$ itself, *but they should scale the same way* if equation (1) is valid. Furthermore, moments of the absolute value of $\delta V$ at non-integer and negative q provide additional useful information about G(u;L',L).

Although the idea of Castaing scaling subtly underlies turbulence theory already, it is instructive to examine the actual kernels G(u;L',L) for popular and reasonable models of the turbulent cascade, and to encourage further turbulence theories to describe themselves with their kernels.

To calculate the cascade kernel G, we use the prescription Muzy, et al. [4] discuss for the case when $\zeta(q)$ does not depend on scale. Then G depends only on u and the ratio L'/L, and the Fourier transform of the G(u;L',L) is equal to

$$\hat{G}(k;L',L) = exp[-ln(L/L')\zeta(-ik)]. \qquad (3)$$

G(u;L',L) is then the inverse Fourier transform. A model with a straight line for the structure function has a delta function kernel (e.g., the kernel for the Kolmogorov's 1941 turbulence model [6] is $\delta[u-(ln(L'/L))/3)]$). Intermittency in the cascade makes $\zeta(q)$ curved, and makes G(u) have finite width which increases with smaller L'.

In the KB 1961 theory [6], $\zeta(q) = (1+\mu/2)(q/3) - \mu q/18$, where $\mu$ is the intermittency parameter. The corresponding kernel for KB 1961 is a Gaussian function whose mean and standard deviation depend on $\mu$ and scale with $\eta \equiv ln(L/L')$:

$$G(u;L',L) = \frac{3}{\sqrt{2\pi\mu\eta}} exp\left\{-\frac{9\left[u + \frac{\eta}{3}\left(1 + \frac{\mu}{2}\right)\right]^2}{2\mu\eta}\right\} . \tag{4}$$

Since u is the logarithm of the scale factor as set up in equation (1), G(u;L',L) for the KB1961 theory is indeed a log-normal distribution of the scale factor.  The scaling of the moments follows from the dependence of G(u;L',L) on L'=$e^{-\eta}$L. G(u;L',L) $\Rightarrow \delta$(u+$\eta$/3) as $\mu \Rightarrow$ 0. Any parabolic $\zeta$(q) has a log-normal kernel; the KB1961 form is the unique parabola which has $\zeta$(3) = 1, and $\zeta$(q)$\Rightarrow$q/3 when $\mu \Rightarrow$0.  Experts dismiss log-normal cascades in incompressible fluid turbulence because the $\zeta$(q) should not have a peak and its slope should be non-negative at all q [6].  In terms of their kernels, the problem with the log-normal models is that their kernels are non-zero (although small), at all positive u. This allows a non-zero probability for arbitrarily large fluctuations at smaller scales to arise from smaller-amplitude fluctuations at larger scales.  These dominate the moments of large positive order, and confuse their scaling. Whether this is also strictly impossible in compressible and/or magnetized fluids such as the solar wind is unclear. Solar wind velocity increments do have $\zeta$(q) with negative slopes in some scale ranges at which the intermittency is very large [7].

Meneveau and Sreenivasan [8] proposed the simple P-model, which avoids these pitfalls. P is a number between 0.5 and 1, typically 0.7, describing the non-uniformity, hence intermittency, in the cascade.  The P-model is a log-*binomial* model. Its $\zeta(q) = 1 - \frac{ln\left(P^{q/3} + (1-P)^{q/3}\right)}{ln(2)}$.  The cascade kernel for the P-model can be derived directly from the description of the model in [8], or by inverting the Fourier transform of the kernel using this $\zeta$(q) in equation (3). Either way, the kernel turns out to be a finite sum of delta-functions, representing the binomial distribution for the number of steps in the cascade. Thus, for L'/L = $2^{-n}$,

$$G(u) = 2^{-n} \sum_0^n \delta\left(u - \frac{m}{3}ln(P) - \frac{n-m}{3}ln(1-P)\right)\frac{n!}{(n-m)!m!} \text{ for the P-model.} \tag{5}$$

As in the KB 1961 model, this G is symmetric about a central maximum value at negative u. When n is not too small, the G for the P-model behaves near its maximum at u = $\eta$log$_2$[P(1-P)]/6 like a KB 1961 log-normal with $\mu$ = {log$_2$[P(1-P)]}-2.

Quite different about the P-model, is that G=0 for all positive u, and u takes exactly n+1 discrete negative values: the most negative is n·ln(1-P)/3 and the closest to zero is n·ln(P)/3. The P-model avoids the pitfalls of the log-normal model by limiting u to negative values. However, it is peculiar that the G is non-zero only at discrete values of u, integer values of n, and that *large negative values of u are absent at all finite scales*.

She and Leveque [9] (hereafter SL) proposed a log-Poisson cascade, which was made more general by Debrulle [10], as a more physical alternative to the log-normal cascade. The cascade kernel corresponding to the original no-parameter SL model and for the Debrulle [10] version with two adjustable parameters is derived below. Both models are log-Poisson, with $\zeta(3) =1$. Debrulle [10] gives the structure function for these models as

$$\zeta(q) = \frac{1-\Delta}{3} q + \frac{\Delta}{1-\beta}\left(1 - \beta^{\frac{q}{3}}\right) \qquad (6)$$

In multifractal terms, $(1-\Delta)/3$ is the minimum scaling exponent and $\Delta/(1-\beta)$ is the co-dimension. The original SL model has $\Delta = \beta = 2/3$. Debrulle [10] speculates that $\Delta$ and $\beta$ may independently take other values between 0 and 1. I find G(u) by straightforwardly inverting the Fourier transform $\hat{G}(k) = exp(-\eta\zeta(-ik))$. After some algebra of interest only to the eccentric, the result is indeed a log-Poisson:

$$G(u) = \sum_{m=0}^{\infty} \delta(u + a\eta + mW)P(m,b\eta) = \sum_{m=0}^{\infty} \delta(u + a\eta + mW)\frac{(b\eta)^m}{m!}e^{-b\eta} \quad , \qquad (7)$$

$$\text{with } a = \frac{1-\Delta}{3} \qquad W = \frac{ln\frac{1}{\beta}}{3} \qquad b = \frac{\Delta}{1-\beta} \qquad \eta = ln(L/L'). \qquad (8)$$

Note that as in the P-model, the G for the log-Poisson models is zero except at discrete values of u. In the log-Poisson models however, the sum extends to $m = \infty$, and arbitrarily large negative u. The log-Poisson models are not symmetric about their maximum values. The minimum scaling exponent, a, makes the largest $u = -a\eta$. The co-dimension, b, determines that the maximum of G occurs at $m = b\eta$. W scales u to m for m > 0. For the original SL model [9], a=1/9, W=ln(1.5)/3, b=2.

In the Debrulle models, $\Delta$ determines where the least negative u will be and has a strong influence on the physicality of the log-Poisson:
1. If $\Delta < 1$, the largest u will be negative, (as in the P-model) and there is a "gap" in the cascade between L and L', corresponding to a minimum scaling exponent > 0. The $\zeta(q)$ for large q approaches a finite positive slope. (If $\Delta = 0$, G(u) consists of one delta function and the K 41 model is recovered.)
2. If $\Delta = 1$, there is no gap, and the minimum scaling exponent is zero. G(0) is non-zero, but G=0 for all u > 0. The $\zeta(q)$ for large q approaches a zero slope. Pagels and Balogh [11] argue that $\zeta(q)$ with zero slope at large q is appropriate for the fast solar wind.

3. If $\Delta > 1$, $a < 0$, so G has non-zero values at a finite number of positive values of u. In this case, the envelope of the kernel is still a Poisson function, but not physical because of those positive values of u.

4. The codimension $b = \Delta/(1-\beta)$ equals the value of m where the kernel takes its largest value. Correspondingly, the u at the largest value of the kernel is $\eta W\Delta/(1-\beta) + \eta(1-\Delta)/3$, so that

$$u_{max}(L') = ln\left(\frac{L'}{L}\right)\left[\Delta\frac{ln(1/\beta)}{3(1-\beta)} + \frac{1-\Delta}{3}\right] = ln\left(\frac{L'}{L}\right)\left[\frac{1}{3} + \frac{\Delta}{3}\left(\frac{ln(1/\beta)}{1-\beta} - 1\right)\right]. \tag{9}$$

The term in square braces $= a + bW$ and corresponds to $d[ln(\sigma_0)]/dln(L) =$ in the Castaing-lognormal model, and $1/3 + \mu/6$ in the K-B 61 model. 2- this term should give the slope of the power spectrum. Again, if $\Delta = 0$, the K41 model is recovered.

These kernels are easily generalized to cascades in which $\zeta(q)$ changes with scale, implying that the parameters such as $\mu$, P, or $\Delta$ and $\beta$ change with scale. (The solar wind has an enormous inertial range, and appears to be much more intermittent at larger scales within the inertial range.)  If parameters, or even the detailed cascade physics, requiring quite different models, vary with scale, G(u;L',L) will not just depend on the ratio of L'/L, but is a convolution of G's over the range from L down to L'. But as long as Castaing scaling applies at each scale, it will apply over the whole inertial range.

For example, if for some L'' such that L>L''>L', the parameters are constant from L down to L'', but have different values from L'' down to L', the net G is a convolution of the G's for the two regimes:

$$G(u;L',L) = \int G(u-u';L',L'')G(u';L'',L)du' \tag{10}$$

In the case where the parameters vary continuously, the Fourier transform of the kernel is

$$\hat{G}(k;L',L) = exp\left\{-\int_{L'}^{L}\zeta_{L''}(-ik)d\,ln L''\right\} = \frac{S(-ik,L')}{S(-ik,L)}. \tag{11}$$

$\zeta(q)$ describes the rate of change of the moments of the kernel G with scale at the observed scale, but the "local" kernel $G(u;e^{-\eta}L,L)$ for small $\eta$ illustrates how the velocity increments (and thus the energy as well) cascades at that scale to a slightly smaller scale. G(u; L',L) contains the cumulation of intermittent effects in the whole turbulent cascade from the scale of energy input down to the observed scale. Since the pdf at small scales is the convolution of that cumulative G with the pdf at the energy scale, it has some of the character of each, notably their mean, width and asymmetry.

What does it mean that the G for the binomial and log-Poisson models is a sum of discrete delta-functions? Is that just an artifact of how the models were constructed? Mathematically, it comes from the factor $e^{-qW}$ in the $\zeta(q)$.  One might think that the turbulent cascade ought to allow a $\delta V$ at some scale to cascade to any smaller $\delta V$ at any smaller scale in the inertial range, not just to certain discrete $\delta V$.  If the parameters of the cascade vary continuously, the delta-functions wash out anyway.

Although the scaling of Castaing kernel functions is completely equivalent to multifractal descriptions of the turbulent cascade, they provide a different kind of insight into models of the cascade and may prove very useful tools in the difficult subject of fluid turbulence.

ACKNOWLEDGEMENTS

This work was supported by NASA grants NAG-58106 and NAG- 510995. My colleagues Leonard Burlaga, Alfred Goldhaber, Timothy Horbury and Ismail Zahed provided useful comments and encouragement.